\documentstyle{article}

\newcommand{\talkauthors}[1]{{\small {\bf #1}}}
\newcommand{\talktitle}[1]{{\small {\bf #1}}}
\newcommand{\address}[1]{{\small {\it #1}}}

\def\thebibliography#1{\centerline{\bf References}
\list
 {\arabic{enumi}.}{\settowidth\labelwidth{[#1]}\leftmargin\labelwidth
 \advance\leftmargin\labelsep
 \usecounter{enumi}}
 \def\newblock{\hskip .11em plus .33em minus .07em}
 \sloppy\clubpenalty4000\widowpenalty4000
 \sfcode`\.=1000\relax}

\newcommand{\be}{\begin{equation}}
\newcommand{\ee}{\end{equation}}
\newcommand{\bea}{\begin{eqnarray}}
\newcommand{\eea}{\end{eqnarray}}

\newcommand{\Int}{\int\limits}
\newcommand{\Li}[2]{{\mbox{Li}}_{#1}\left(#2\right)}
\newcommand{\sgn}[1]{\mbox{sgn}\left(#1\right)}

\begin{document}
\vspace*{25mm}
\begin{center}
 \talktitle{SOME EXACT RESULTS FOR TWO-LOOP DIAGRAMS \\
             WITH THREE AND FOUR EXTERNAL LINES
\footnote{ To appear in Proc. VII Intern. Workshop
"High Energy Physics and Quantum Field Theory"
(Sochi, Russia, October 1992), {\em Yad.Fiz.},
v.56, No.11, p.172-179 (November 1993).}
}\\ [5mm]
\talkauthors{N.\ I.\ USSYUKINA \ and
 A.\ I.\  DAVYDYCHEV
 }
\vskip 4mm
\address{NUCLEAR PHYSICS INSTITUTE, MOSCOW STATE UNIVERSITY \\
          RUSSIA} \\
\end{center}
\vspace{4mm}

{\small
Evaluation of three- and four-point diagrams with massless
internal particles and arbitrary external momenta is considered.
Exact results for some two-loop diagrams (planar and non-planar
three-point contributions and the ``double box'' diagram)
are obtained in terms of polylogarithms.
}

\vspace{5mm}

{\bf 1.} Recent development of modern accelerators requires information
about higher-order loop corrections to elementary particle processes.
Some of the interesting problems (corrections to multijet processes,
Bhabha scattering, etc.) are connected with the evaluation of
two-loop three- and four-point diagrams.
The case of massless internal propagators is of special interest,
because often we are confronted either with really massless particles
(photon, gluon) or with particles whose masses can be neglected
in high-energy processes (electron, light quarks, etc.). For example,
in a number of publications \cite{OW,BS,GT(W)} (see also \cite{ELOP}
and references therein) the asymptotic behaviour of ladder diagrams
has been examined for high energies and momentum transfer in leading
logarithmic approximation.

The present paper is devoted to some methods which enable one to
obtain exact expressions for some types of Feynman diagrams
with massless internal particles and arbitrary external
momenta (with three and four external particles).
The used approach involves the following tools: (i) the Feynman
parametric representation, (ii) the ``uniqueness'' conditions (see, e.g.,
in \cite{Us'83}), (iii) Mellin--Barnes contour integrals and
(iv) Fourier transform to the coordinate space.
We shall consider only scalar diagrams (corresponding to
massless $\phi^3$ theory), because expressions occurring in realistic
calculations can be reduced to such scalar integrals (see, e.g.,
\cite{PV}).

The remainder of the paper is organized as follows. In Section 2 we
present the main steps of the approach via examples of one-loop
triangle and box diagrams. In Section 3 we apply this technique to
obtain expressions for two-loop three- and four-point
ladder (planar) diagrams. Section 4 discusses the evaluation of
some non-planar graphs. Section 5 formulates and discusses the
main results.

\vspace{0.4cm}

{\bf 2.} Let us start by some definitions and results for massless
triangle diagrams (see Fig.~1).
\begin{figure}[bth]
\centering
\setlength {\unitlength}{0.7mm}
\begin{picture}(150,64)(0,0)
\put (50,10) {\line(3,4){24}}
\put (50,10) {\line(1,0){48}}
\put (98,10) {\line(-3,4){24}}
\put (74,42) {\line(0,1){14}}
\put (50,10) {\line(-3,-2){14}}
\put (98,10) {\line(3,-2){14}}
\put (50,10) {\circle*{2}}
\put (98,10) {\circle*{2}}
\put (74,42) {\circle*{2}}
\put (32,10) {\makebox(0,0)[bl]{\large $p_2$}}
\put (107,10) {\makebox(0,0)[bl]{\large $p_1$}}
\put (80,50) {\makebox(0,0)[bl]{\large $p_3$}}
\put (73,3) {\makebox(0,0)[bl]{\large $\nu_3$}}
\put (91,30) {\makebox(0,0)[bl]{\large $\nu_2$}}
\put (50,30) {\makebox(0,0)[bl]{\large $\nu_1$}}
\end{picture}
\caption{}

\end{figure}
Here and below we shall consider all external momenta to be
ingoing ($ p_1 + p_2 + p_3 =0$). The corresponding Feynman integral is
\be
\label{defJ}
J(n; \; \nu_1  ,\nu_2  ,\nu_3  ) \equiv \int
 \frac{d^n k}{ ((q_1 +k )^2)^{\nu_1}  ((q_2 +k )^2)^{\nu_2}
      ((q_3 +k )^2)^{\nu_3} }
\ee
where $q_3-q_2=p_1, q_1-q_3=p_2, q_2-q_1=p_3$, and $n$ is the
space-time dimension. The usual ``causal'' prescription for
singularities in pseudo-Euclidean momentum space is understood,
\be
\label{i0}
\vspace{-1mm}
((q+k)^2 )^{-\nu} \leftrightarrow ((q+k)^2 +i0)^{-\nu}  .
\ee

When the powers of denominators and $n$ are related by
$\nu_1+ \nu_2+ \nu_3 = n$, a very simple result can be obtained
for such a ``unique'' triangle \cite{D'E,Vas}:
\be
\label{uniq}
\left. \frac{}{}
J(n; \; \nu_1,\nu_2,\nu_3) \right|_{\Sigma \nu_i = n}
=  \pi^{n/2} \; i^{1-n} \prod_{i=1}^{3}
\frac{\Gamma ( n/2 - \nu_i)}{\Gamma (\nu_i)} (p_i^2)^{\nu_i -n/2}.
\ee
If $\nu_1+ \nu_2+ \nu_3 =n-1$,
then the following identity \cite{Us'83} (see also \cite{BU'83,Kaz})
holds for such ``semi-unique'' triangles:
\\ \vspace{0.1cm} \\
\begin{math}
\left\{ \left.   \frac{}{}
\nu_1 J(n; \; \nu_1 +1,\nu_2,\nu_3)
+ \nu_2 J(n; \; \nu_1 ,\nu_2 +1,\nu_3)
+ \nu_3 J(n; \; \nu_1 ,\nu_2,\nu_3 +1) \right\}
\right|_{\Sigma \nu_i = n-2}
\end{math}
\be
\label{uniq2}
\hspace{2cm}   =  \pi^{n/2} \; i^{1-n} \prod_{i=1}^{3}
\frac{\Gamma (n/2- \nu_i -1)}{\Gamma (\nu_i)} (p_i^2)^{\nu_i-n/2+1} .
\ee
We shall use so-called ``uniqueness'' conditions (\ref{uniq}) and
(\ref{uniq2}) below, when evaluating the two-loop planar diagram.

In applications the case $\nu_1= \nu_2= \nu_3 =1, \; n=4$ is
important. Let us denote
\be
\label{defC1}
C^{(1)}(p_1^2, p_2^2, p_3^2) \equiv  J(4; 1, 1, 1) .
\ee
Note that (\ref{defC1}) can be represented in terms
of a two-fold Mellin--Barnes integral \cite{Us'75}
\be
\label{MBC1}
\vspace{-1mm}
C^{(1)} = \frac{i \pi^2}{p_3^2}
\frac{1}{(2\pi i)^2}
\Int_{-i\infty}^{i\infty} \Int_{-i\infty}^{i\infty}
du\; dv\; x^u\; y^v \;
\Gamma^2 (-u) \; \Gamma^2 (-v) \; \Gamma^2 (1+u+v) ,
\ee
where
\be
\label{xy}
\vspace{-1mm}
 x \equiv \frac{p_1^2}{p_3^2} \hspace{0.5cm}\mbox{   ,   }
\hspace{0.5cm} y \equiv \frac{p_2^2}{p_3^2} \; \; ,
\ee
and (here and below) the integration contours are chosen so as to separate
the  ``right'' and ``left'' series of poles of gamma functions in the
integrand (see, e.g., \cite{Bailey}). Analogous Mellin--Barnes
representation for the case of arbitrary values of $n$ and $\nu_i$
can be found in \cite{BD'91,JPA}.

Using Feynman parameteric representation
(or (\ref{MBC1})) yields
\be
\label{defPhi}
C^{(1)}(p_1^2, p_2^2, p_3^2) =
\frac{i \pi^2}{p_3^2}  \; \Phi^{(1)} (x,y)
\ee
with
\be
\label{intPhi}
\vspace{-1mm}
\Phi^{(1)} (x,y) = - \Int_0^1
 \frac{d\xi}{y \xi^2 + (1-x-y) \xi + x}
 \left( \ln{\frac{y}{x}} + 2 \ln{\xi} \right) \; .
\ee
The integral (\ref{intPhi}) can be evaluated in terms of
dilogarithms $\Li{2}{z}$ (see, e.g., \cite{Lewin}),
\be
\label{Phi}
\Phi^{(1)} (x,y) = \frac{1}{\lambda} \left\{ \frac{}{}
2 \left( \Li{2}{-\rho x} + \Li{2}{-\rho y} \right)
+ \ln(\rho x) \ln(\rho y)
+ \ln{\frac{y}{x}} \ln{\frac{1+\rho y}{1+\rho x}} + \frac{\pi^2}{3}
\right\}
\ee
with
\be
\label{lambda-rho}
\lambda(x,y) \equiv \sqrt{(1-x-y)^2 - 4 x y} \; \; ,
\hspace{8mm}
\rho(x,y) \equiv \frac{2}{1-x-y+\lambda} \; .
\ee
Note that representations of such type are well known for triangle
diagrams (see, e.g., \cite{'tHV'79}). By simple dilogarithm
transformations, formula (\ref{Phi}) can be turned into the result
obtained in \cite{JPA}. In the paper \cite{INLO11} some problems
of analytic continuation of the function $\Phi^{(1)} (x,y)$ (\ref{Phi})
in the region of positive $x$ and $y$ were examined. If we consider,
for example, negative values of both variables $x$ and $y$, we
should take into account the prescription (\ref{i0}). This requires
the following substitutions in (\ref{Phi}):
\be
\label{neg-xy}
\ln(\rho x) \rightarrow \ln(-\rho x) + i \pi \sigma \; ; \hspace{1cm}
\ln(\rho y) \rightarrow \ln(-\rho y) + i \pi \sigma \; ,
\ee
where $\sigma \equiv \sgn{p_3^2}$.

Let us consider now a four-point ``box'' diagram in Fig.~2.
\begin{figure}[bth]
\centering
\setlength {\unitlength}{0.7mm}
\begin{picture}(150,60)(0,0)
\put (40,10) {\line(1,0){80}}
\put (40,50) {\line(1,0){80}}
\put (60,10) {\line(0,1){40}}
\put (100,10) {\line(0,1){40}}
\put (60,10) {\circle*{2}}
\put (60,50) {\circle*{2}}
\put (100,10) {\circle*{2}}
\put (100,50) {\circle*{2}}
\put (32,10) {\makebox(0,0)[bl]{\large $k_1$}}
\put (32,48) {\makebox(0,0)[bl]{\large $k_2$}}
\put (124,48) {\makebox(0,0)[bl]{\large $k_3$}}
\put (124,10) {\makebox(0,0)[bl]{\large $k_4$}}
\end{picture}
\caption{}

\end{figure}
All external momenta are taken to be ingoing ($ k_1+k_2+k_3+k_4=0 $).
We shall denote this diagram as
$D^{(1)}(k_1^2, k_2^2, k_3^2, k_4^2, s, t)$ with
\be
\label{st}
s \equiv (k_1 + k_2)^2 \; , \; \; \; t \equiv (k_2+k_3)^2 .
\ee
Then, by use of Feynman parametrization, uniqueness condition (\ref{uniq})
and Mellin--Barnes representation, it is possible to show \cite{UD} that
\be
\label{MBD1}
\vspace{-1mm}
D^{(1)} =
\frac{i \pi^2}{s \; t}
\frac{1}{(2\pi i)^2}
\Int_{-i\infty}^{i\infty} \Int_{-i\infty}^{i\infty}
du\; dv\; X^u\; Y^v \;
\Gamma^2 (-u) \; \Gamma^2 (-v) \; \Gamma^2 (1+u+v) \; ,
\ee
where
\be
\label{XY}
X \equiv \frac{k_1^2 k_3^2}{s \; t} \; \; , \;\;\;\;
Y \equiv \frac{k_2^2 k_4^2}{s \; t} .
\ee
Comparison with the representation (\ref{MBC1}) gives
\be
\label{D1C1}
D^{(1)}(k_1^2, k_2^2, k_3^2, k_4^2, s, t)
= C^{(1)} (k_1^2 k_3^2, k_2^2 k_4^2, s t) ,
\vspace{-1mm}
\ee
or
\be
\label{D1Phi}
\vspace{-1mm}
D^{(1)}(k_1^2, k_2^2, k_3^2, k_4^2, s, t)
= \frac{i \pi^2}{s \; t} \; \Phi^{(1)} (X,Y) ,
\ee
with the same function $\Phi^{(1)}$ as for triangle diagram. Thus,
the result obtained for the box diagram contains only two
dilogarithms. For negative $x$ and $y$ (this corresponds, e.g.,
to a physically interesting case when $s$ and $k_i^2$ are positive while
$t$ is negative), it is necessary to use a prescription of the type of
(\ref{neg-xy}) (for details see in \cite{DNS}).
We note a useful fact of ``pairing'' of variables in the
four-point function (\ref{D1C1}) allowing reduction to
the three-point function. Below we shall see that an analogous
property also occurs for two-loop ladder diagrams.

\vspace{0.4cm}

{\bf 3.} In this section we shall consider the main steps of
evaluating two-loop three- and four-point ladder graphs. A more
detailed information can be found in \cite{UD}.

Let us consider first a three-point two-loop ladder (planar)
diagram shown in Fig.~3 ($p_1+p_2+p_3=0$).
\begin{figure}[bth]
\centering
\setlength {\unitlength}{0.7mm}
\begin{picture}(150,58)(0,0)
\put (30,30) {\line(1,0){20}}
\put (50,30) {\line(3,1){80}}
\put (50,30) {\line(3,-1){80}}
\put (80,20) {\line(0,1){20}}
\put (110,10) {\line(0,1){40}}
\put (50,30) {\circle*{2}}
\put (80,20) {\circle*{2}}
\put (80,40) {\circle*{2}}
\put (110,10) {\circle*{2}}
\put (110,50) {\circle*{2}}
\put (22,30) {\makebox(0,0)[bl]{\large $p_3$}}
\put (134,52) {\makebox(0,0)[bl]{\large $p_1$}}
\put (134,7) {\makebox(0,0)[bl]{\large $p_2$}}
\put (56,40) {\makebox(0,0)[bl]{\large $1+\delta_1$}}
\put (56,14) {\makebox(0,0)[bl]{\large $1+\delta_2$}}
\put (86,52) {\makebox(0,0)[bl]{\large $1+\delta_2$}}
\put (86,6) {\makebox(0,0)[bl]{\large $1+\delta_1$}}
\put (83,29) {\makebox(0,0)[bl]{\large $1+\delta_3$}}
\put (113,29) {\makebox(0,0)[bl]{\large $1+\delta_3$}}
\end{picture}
\caption{}

\end{figure}
The Feynman integral corresponding to this diagram (with unit powers
of denominators) can be written as
\be
\label{defC2}
\vspace{-1mm}
C^{(2)}(p_1^2, p_2^2, p_3^2) = \int
\frac{d^4 r}{r^2 \; (p_1+r)^2 \; (p_2-r)^2} \;
C^{(1)}((p_1+r)^2, (p_2-r)^2, p_3^2) ,
\ee
where $C^{(1)}$ is the one-loop function (\ref{defC1}).

To calculate $C^{(2)}$, it is convenient to use the ``uniqueness''
method, by analogy with the paper \cite{BU'83} (where
propagator-type ladder diagrams have been examined). To do this,
let us consider a special analytic regularization of this
diagram (see Fig.3), where we replace unit powers of denominators
by $(1+ \delta_i)$ , provided that $\delta_1 + \delta_2 +\delta_3=0$.
Applying relations (\ref{uniq}) and (\ref{uniq2}) to this
regularized diagram gives the following result (at $n=4$):
\bea
\label{uniqC2}
\vspace{-1mm}
\frac{i \pi^2}{(p_3^2)^{1-\delta_3}}
\prod \frac{\Gamma(1- \delta_i)}{\Gamma(1+ \delta_i)}
\left\{ \frac{1}{\delta_1 \delta_2} J(4; \; 1, 1, 1+\delta_3) \right.
\hspace{3.5cm} \nonumber \\
\left.
+\frac{1}{\delta_1 \delta_3} (p_1^2)^{\delta_1} J(4; \; 1, 1, 1-\delta_2)
+\frac{1}{\delta_2 \delta_3} (p_2^2)^{\delta_2} J(4; \; 1, 1, 1-\delta_1)
\right\} ,
\eea
where one-loop integrals $J$ are defined in (\ref{defJ}).
Note that integrals on the r.h.s. of (\ref{uniqC2}) can be
transformed by use of the formulae
\be
\label{trJ}
J(4; \; 1, 1, 1+\delta)
= (p_1^2)^{-\delta} J(4; \; 1+\delta, 1-\delta, 1)
= (p_2^2)^{-\delta} J(4; \; 1-\delta, 1+\delta, 1) .
\ee
These relations can also be obtained from (\ref{uniq}) and
(\ref{uniq2}).

By use of (\ref{trJ}) and Feynman parameters, after
some transformations one can obtain the following representation:
\be
\label{intJdel}
\vspace{-1mm}
J(4; 1, 1, 1+\delta) = \frac{i \pi^2}{(p_3^2)^{1+\delta}}
 \; \frac{1}{\delta} \Int_0^1 d \xi \;
\frac{(y \xi)^{-\delta} - (x/ \xi)^{-\delta}}
     {y \xi^2 + (1-x-y) \xi + x} \; .
\vspace{-1mm}
\ee
Inserting (\ref{intJdel}) into (\ref{C2Phi}) we get (as
$\delta_i \rightarrow 0, \; \sum \delta_i = 0 $) :
\be
\label{C2Phi}
\vspace{-1mm}
C^{(2)} (p_1^2, p_2^2, p_3^2) = \left( \frac{i \pi^2}{p_3^2} \right)^2
   \; \Phi^{(2)} (x,y)  \; ,
\vspace{-1mm}
\ee
where
\be
\label{Phi2int}
\vspace{-2mm}
\Phi^{(2)} (x,y) = -\frac{1}{2}
\Int_0^1 \frac{d \xi \; \ln{\xi}}{y \xi^2 + (1-x-y) \xi + x}
 \left( \ln{\frac{y}{x}} + \ln{\xi} \right)
 \left( \ln{\frac{y}{x}} + 2 \ln{\xi} \right) .
\vspace{-1mm}
\ee
This integral can be easily calculated in terms of
polylogarithms $\Li{N}{z}$ (see \cite{Lewin}),
\be
\label{LiN}
\vspace{-1mm}
\Li{N}{z} = \frac{(-1)^N}{(N-1)!} \Int_0^1 d \xi \;
\frac{\ln^{N-1} \xi}{\xi -z^{-1}} .
\vspace{-1mm}
\ee
So, we arrive at the following result for the two-loop three-point
diagram of Fig.~3:
\bea
\label{Phi2Li}
\Phi^{(2)} (x,y) =
 \frac{1}{\lambda}
\left\{6 \left( \Li{4}{-\rho x} + \Li{4}{-\rho y} \right)
    + 3 \ln \frac{y}{x} \left( \Li{3}{-\rho x} - \Li{3}{-\rho y} \right)
\right.
\nonumber \\ \hspace{2cm}
 + \frac{1}{2} \ln^2 \frac{y}{x}
        \left( \Li{2}{-\rho x} + \Li{2}{-\rho y} \right)
 + \frac{1}{4} \ln^2 (\rho x) \ln^2 (\rho y)  \hspace{1cm}
\nonumber \\ \hspace{2cm}  \left.
 + \frac{\pi^2}{2} \ln (\rho x) \ln (\rho y)
 + \frac{\pi^2}{12} \ln^2 \frac{y}{x} +\frac{7 \pi^4}{60} \right\},
\hspace{2cm}
\eea
where $\lambda(x,y)$ and $\rho(x,y)$ are defined in (\ref{lambda-rho}).
For negative values of $x$ and $y$,
one has to use (\ref{neg-xy}).

Let us consider now the two-loop four-point diagram (``double box'')
presented in Fig.~4.
\begin{figure}[bth]
\centering
\setlength {\unitlength}{0.7mm}
\begin{picture}(150,50)(0,0)
\put (30,10) {\line(1,0){90}}
\put (30,40) {\line(1,0){90}}
\put (45,10) {\line(0,1){30}}
\put (75,10) {\line(0,1){30}}
\put (105,10) {\line(0,1){30}}
\put (45,10) {\circle*{2}}
\put (45,40) {\circle*{2}}
\put (75,10) {\circle*{2}}
\put (75,40) {\circle*{2}}
\put (105,10) {\circle*{2}}
\put (105,40) {\circle*{2}}
\put (22,10) {\makebox(0,0)[bl]{\large $k_1$}}
\put (22,38) {\makebox(0,0)[bl]{\large $k_2$}}
\put (124,38) {\makebox(0,0)[bl]{\large $k_3$}}
\put (124,10) {\makebox(0,0)[bl]{\large $k_4$}}
\end{picture}
\caption{}

\end{figure}
All notations correspond to the one-loop case (in particular, $s$ and $t$
are defined by (\ref{st})). The corresponding integral can be
represented as
\bea
\label{defD2}
\vspace{-1mm}
D^{(2)} (k_1^2, k_2^2, k_3^2, k_4^2, s, t)
= \int \frac{d^4 r}{r^2 (k_3+r)^2 (k_4-r)^2}
\hspace{3cm} \nonumber \\
\times D^{(1)} (k_1^2, k_2^2, (k_3+r)^2, (k_4-r)^2, s, (k_2+k_3+r)^2 ),
\eea
where $D^{(1)}$ is one-loop function (see Fig.~2).
Note that the $s$-channel of the diagram in Fig.4 corresponds to
a ``horizontal double box'' with initial particles momenta $k_1$
and $k_2$. On the other hand, the $t$-channel corresponds to
a ``vertical double box'' with initial momenta $k_2$ and $k_3$.

By using the same technique as for one-loop functions, one can
construct four-fold Mellin--Barnes representations for $C^{(2)}$
and $D^{(2)}$ (see \cite{UD}). Comparing these representations
we find
\be
\label{D2C2}
D^{(2)} (k_1^2, k_2^2, k_3^2, k_4^2, s, t)
= t \; C^{(2)} (k_1^2 \; k_3^2 ,\; k_2^2 \; k_4^2 , \; s \; t) ,
\vspace{-1mm}
\ee
or
\be
\label{D2Phi}
\vspace{-2mm}
D^{(2)} (k_1^2, k_2^2, k_3^2, k_4^2, s, t)
= \frac{(i \pi^2)^2}{s^2 \; t} \; \Phi^{(2)} (X,Y)
\ee
with $X$ and $Y$ defined by (\ref{XY}).
We see that in two-loop case we also have obtained a ``pairing'' of
four-point function arguments. As a result, $D^{(2)}$ can be reduced
to three-point function $C^{(2)}$ (\ref{C2Phi})-(\ref{Phi2Li}) (where
we should make the substitutions $p_3^2 \rightarrow s t, \; x \rightarrow X,
\; y \rightarrow Y$ ; see (\ref{XY})). Thus, formula (\ref{Phi2Li}) (combined
with (\ref{D2Phi}) and (\ref{XY})) yields a representation of the
``double box'' $D^{(2)}$ (Fig.4) in terms of polylogarithms (\ref{LiN}).

\vspace{0.4cm}

{\bf 4.} In this section we shall examine an interesting example of
non-planar graphs: two-loop three-point crossed diagram
$\widetilde{C}^{(2)}(p_1^2, p_2^2, p_3^2)$ shown in Fig.~5.
\begin{figure}[b]
\centering
\setlength {\unitlength}{0.7mm}
\begin{picture}(150,58)(0,0)
\put (30,30) {\line(1,0){20}}
\put (50,30) {\line(3,1){80}}
\put (50,30) {\line(3,-1){80}}
\put (80,20) {\line(1,1){30}}
\put (80,40) {\line(1,-1){30}}
\put (50,30) {\circle*{2}}
\put (80,20) {\circle*{2}}
\put (80,40) {\circle*{2}}
\put (110,50) {\circle*{2}}
\put (110,10) {\circle*{2}}
\put (22,30) {\makebox(0,0)[bl]{\large $p_3$}}
\put (134,52) {\makebox(0,0)[bl]{\large $p_1$}}
\put (134,7) {\makebox(0,0)[bl]{\large $p_2$}}
\put (47,34) {\makebox(0,0)[bl]{\large $A_3$}}
\put (107,54) {\makebox(0,0)[bl]{\large $A_1$}}
\put (107,1) {\makebox(0,0)[bl]{\large $A_2$}}
\put (77,44) {\makebox(0,0)[bl]{\large $V_1$}}
\put (77,11) {\makebox(0,0)[bl]{\large $V_2$}}
\end{picture}
\caption{}

\end{figure}
We see that each single loop of this diagram corresponds to
four-point function $D^{(1)}$.

To evaluate the diagram in Fig.~5, it is convenient to use
Fourier transform to coordinate space. It is easy to show
that the ``topology'' of the diagram remains the same in the
$x$-space (see Fig.~5): there are three points $A_1$, $A_2$,
$A_3$ (associated with external vertices), and there are
two points $V_1$ and $V_2$ connected by lines with each of
$A_i$. Note that in the $x$-space both integrations (with
respect to the positions of $V_1$ and $V_2$) are independent.
Therefore, the crossed diagram in Fig.~5 factorizes and,
using the $x$-space ``uniqueness'' condition (see, e.g., in
\cite{Us'83}) and returning to the momentum space, we find a
simple relation:
\be
\label{Cc2C1}
\vspace{-1mm}
\widetilde{C}^{(2)}(p_1^2, p_2^2, p_3^2)
= \left( C^{(1)}(p_1^2, p_2^2, p_3^2) \right) ^2 \; ,
\ee
or (see (\ref{defPhi}))
\be
\label{Cc2Phi}
\vspace{-1mm}
\widetilde{C}^{(2)}(p_1^2, p_2^2, p_3^2)
= \left( \frac{i \pi^2}{p_3^2} \right) ^2
  \left( \Phi^{(1)} (x,y) \right) ^2
\ee
with $\Phi^{(1)} (x,y)$ defined by (\ref{intPhi})-(\ref{Phi}).
So, the result for the crossed diagram in Fig.~5 is expressed
in terms of the product of dilogarithms.

We can also use the formulae (\ref{Cc2C1})-(\ref{Cc2Phi}) to
check the known result for three-loop crossed propagator-type
diagram (see Fig.~6),
\begin{figure}[t]
\centering
\setlength {\unitlength}{0.7mm}
\begin{picture}(150,50)(0,0)
\put (15,25) {\line(1,0){30}}
\put (45,25) {\line(1,1){15}}
\put (45,25) {\line(1,-1){15}}
\put (60,10) {\line(1,0){30}}
\put (60,40) {\line(1,0){30}}
\put (60,10) {\line(1,1){30}}
\put (60,40) {\line(1,-1){30}}
\put (105,25) {\line(-1,1){15}}
\put (105,25) {\line(-1,-1){15}}
\put (105,25) {\line(1,0){30}}
\put (45,25) {\circle*{2}}
\put (105,25) {\circle*{2}}
\put (60,10) {\circle*{2}}
\put (60,40) {\circle*{2}}
\put (90,10) {\circle*{2}}
\put (90,40) {\circle*{2}}
\put (20,30) {\vector(1,0){13}}
\put (25,32) {\makebox(0,0)[bl]{\large $k$}}
\end{picture}
\caption{}

\end{figure}
\be
\widetilde{B}^{(3)} (k^2) = \int \frac{d^4 r}{r^2 \; (k+r)^2} \;
  \widetilde{C}^{(2)} (r^2, (k+r)^2, k^2) .
\ee

Representing $\widetilde{C}^{(2)}$ in terms of a product of
one-loop functions (\ref{Cc2C1})-(\ref{Cc2Phi}) and using
parametric representation (\ref{intPhi}) for $\Phi^{(1)}$
functions (this yields integrations over the parameters
$\xi$ and $\xi'$), we obtain that the integral over $r$
corresponds to a box diagram (Fig.~2) with ingoing momenta
$k_1 = (1-\xi)^{-1} k$, $k_2 = -\xi (1-\xi)^{-1} k$,
$k_3 = \xi' (1-\xi')^{-1} k$, $k_4 = -(1-\xi')^{-1} k$.
The logarithms $\ln((k+r)^2/r^2)$ occurring in the numerator
can be transformed into derivatives with respect to the
powers of denominators by
\be
\label{ln-der}
\frac{1}{r^2 \; (k+r)^2} \ln^j \left( \frac{(k+r)^2}{r^2} \right)
 = \left( \frac{\partial^j}{\partial \delta^j}
   \left. \frac{1}{\left( r^2 \right) ^{1+\delta} \;
                   \left( (k+r)^2 \right) ^{1-\delta}} \right)
    \right |_{\delta = 0} . \;
\ee
By use of the same technique as in Section~2 the box integral
with shifted powers of denominators $(1+\delta, \; 1-\delta)$
can be reduced to the three-point function (\ref{trJ}). Using
the parametric representation (\ref{intJdel}) and making evident
substitutions of variables, we arrive at the well-known result
(see, e.g., \cite{CKT}):
\be
\label{zeta5}
\vspace{-1mm}
\widetilde{B}^{(3)} (k^2)
 = -\frac{1}{6} \; \frac{(i \pi^2)^3}{(k^2)^2}
  \Int_0^1 \frac{d \xi}{(1-\xi)^2} \ln^5 \xi
 = \frac{(i \pi^2)^3}{(k^2)^2} \; 20 \; \zeta(5) .
\vspace{-1mm}
\ee
It should be noted that by using the representation (\ref{intJdel})
the same result (\ref{zeta5}) can be obtained also for
the planar three-loop propagator-type diagram.

\vspace{0.4cm}

{\bf 5.} In the present paper we considered (via two-loop
examples) the evaluation of planar and non-planar diagrams with
three and four external lines with arbitrary momenta. We
used Feynman parametrization, ``uniqueness'' conditions
(\ref{uniq})-(\ref{uniq2}), Mellin--Barnes contour integrals
and Fourier transformation. For the ladder diagrams,
it is shown that the corresponding four-point functions can
be reduced to three-point ones (see (\ref{D1C1}) and (\ref{D2C2})).
Note that analogous formulae also occur for ladder diagrams
with arbitrary number of rungs.

The results (\ref{Phi}) and (\ref{Phi2Li}) are presented in terms of
polylogarithms with simple arguments (it is convenient to use
such expressions in realistic calculations). Note that applying
relations (\ref{uniq}) and (\ref{uniq2}) makes it possible to
consider ladder diagrams with any numbers of loops.
These results also can be expressed in terms of polylogarithms
(\ref{LiN}). In the general case of $L$-loop ladder diagram, the
highest order of occurring polylogarithms is equal to $2L$.

It should be noted that, if external momenta vanish, we get infrared
(on-shell) singularities. For example, in the papers
\cite{Gons,vN,KL} the diagrams of Fig.~3 and Fig.~5 have been
considered at $p_1^2 = p_2^2 = 0$ by use of dimensional regularization,
and singularities have appeared as the poles in
$\varepsilon = (4-n)/2 $. The leading singularities of $C^{(2)}$
and $\widetilde{C}^{(2)}$ are of the order of $1/\varepsilon^4$
(note that the coefficient at $1/\varepsilon^4$ in $\widetilde{C}^{(2)}$
is four times as large than one in $C^{(2)}$).
In our (four-dimensional) approach in this case we can put
$p_{1,2}^2 = \mu^2$ ( $\mu \rightarrow 0$), and the singularities
will be manifested as powers of $\ln \mu$. It is easy to see
from (\ref{Cc2Phi}), (\ref{Phi}) and (\ref{C2Phi}), (\ref{Phi2Li})
that in this case the leading singularities of $\widetilde{C}^{(2)}$
and $C^{(2)}$ are given by $\ln^2 (\rho x) \ln^2 (\rho y) \sim
\ln^4 \mu$, the ratio of coefficients being also equal to four.
On the other hand, there is a problem how to relate more soft
logarithmic  and $1/\varepsilon$ singularities.

\vspace{0.3cm}

{\bf Acknowledgements}. One of the authors (A.D.) is grateful to
F.A.Berends and  W.L. van Neerven  for drawing his attention to
importance of the ``double box'' diagram. We would like to thank
D.J.Broadhurst, D.Kreimer, U.Nierste and F.V.Tkachov for useful
discussions.

\vspace{0.5cm}

\end{document}